\documentclass[aps,preprint]{revtex4}%
\usepackage{amsfonts}
\usepackage{amsmath}
\usepackage{amssymb}
\usepackage{graphicx}%
\begin{document}
\preprint{CAMS/03-07}
\title[Gravity with Complex Vierbein]{$SL(2,\mathbb{C})$ Gravity with Complex Vierbein and its Noncommutative Extension}
\author{Ali H. Chamseddine\thanks{Electronic mail: chams@aub.edu.lb}}
\affiliation{Center for Advanced Mathematical Sciences (CAMS) and Physics Department,
American University of Beirut}
\keywords{complex gravity, bigravity, noncommutative geometry}
\pacs{PACS 04.20.Fy}
\pacs{PACS 02.40.Gh}
\pacs{PACS 04.20.Fy}
\pacs{PACS 02.40.Gh}
\pacs{PACS 04.20.Fy}
\pacs{PACS 02.40.Gh}
\pacs{PACS 04.20.Fy}
\pacs{PACS 02.40.Gh}
\pacs{PACS 04.20.Fy}
\pacs{PACS 02.40.Gh}
\pacs{PACS 04.20.Fy}
\pacs{PACS 02.40.Gh}
\pacs{PACS 04.20.Fy}
\pacs{PACS 02.40.Gh}
\pacs{PACS 04.20.Fy}
\pacs{PACS 02.40.Gh}

\begin{abstract}
We show that it is possible to formulate gravity with a complex vierbein based
on $SL(2,\mathbb{C})$ gauge invariance. The proposed action is a four-form
where the metric is not introduced but results as a function of the complex
vierbein. This formulation is based on the first order formalism. The novel
feature here is that integration of the spin-connection gauge field gives rise
to kinetic terms for a massless graviton, a massive graviton with the
Fierz-Pauli mass term, and a scalar field. The resulting theory is equivalent
to bigravity. We then show that by extending the gauge group to
$GL(2,\mathbb{C)}$ the formalism can be easily generalized to apply to a
noncommutative space with the star product. We give the deformed action and
derive the Seiberg-Witten map for the complex vierbein and gauge fields.

\end{abstract}
\volumeyear{year}
\volumenumber{number}
\issuenumber{number}
\eid{identifier}
\startpage{1}
\endpage{2}
\maketitle





\section{Introduction}

\noindent The general theory of relativity can be formulated either as a
geometrical theory in terms of a metric tensor over the space-time manifold,
or in terms of a vierbein and a spin-connection of the local Lorentz algebra
\cite{utiyama}, \cite{kibble}. Both formulations lead to equivalent results as
far as the dynamics of the graviton is concerned. The second approach is more
appropriate to couple to spinors \cite{weyl}, \cite{cartan}, \cite{penrose}.
There were attempts to unify gravitation with other interactions, notably the
Kaluza-Klein approach of compactifying higher dimensional theories, and the
Einstein- Strauss-Schr\"{o}dinger \cite{Einstein}, \cite{schrodinger} approach
of considering a Hermitian metric tensor and interpreting the antisymmetric
field as that of the Maxwell field strength. The advantages and disadvantages
of the Kaluza-Klein approach are well known while the uses of complex
space-time metric are less familiar \cite{Moffat}, \cite{siegel}. It is now
well known that the antisymmetric part of the Hermitian metric cannot be
interpreted as the photon field strength but rather as an antisymmetric tensor
where the theory is consistent only if the field is massive \cite{deser}.
Recently, a formulation of complex gravity using the idea of gauging the
unitary algebra $U(2,2)$ was made using a complex vierbein \cite{chams}. This
was shown to give an action with many desirable properties, the main
disadvantage is that the density formed from the complex vierbein is not
unique. As one of the motivations for introducing a complex metric is to
deform general relativity for a special noncommutative space with a star
product, it is necessary to require the full action to be invariant under both
the star product and the group transformations. The easiest way to implement
this requirement is to construct the action to be a trace of a four-form,
insuring that it is a gauge invariant density. It turned out that in this case
it is not easy to obtain a simple action satisfying these properties. By using
a constrained gauge group $U(2,2)$ the construction becomes possible, but only
for conformal gravity, not Einstein gravity. Another disadvantage is that it
was necessary to use the Seiberg-Witten map \cite{SW}, \cite{wess} in order to
solve the noncommutative constraints, resulting in complicated expressions.

It is therefore important to have a gauge invariant formulation of deformed
complex gravity where the action is written as a four-form. To do this we must
first succeed in formulating complex gravity without introducing apriori a
metric tensor. Taking a close look at the $SL(2,\mathbb{C})$ formulation of
gravity \cite{salam} one notes that the following steps are needed. First an
$SL(2,\mathbb{C})$ gauge field is introduced, the spin-connection,
\[
\omega=dx^{\mu}\omega_{\mu}=\frac{1}{4}dx^{\mu}\omega_{\mu ab}\gamma^{ab},
\]
where $\gamma^{ab}$ is the antisymmetrized product of Dirac gamma
matrices\footnote{We adopt the notation of reference \cite{van} for the Dirac
gamma matrices. In particular $\left\{  \gamma_{a},\gamma_{b}\right\}
=2\delta_{ab}$, $\gamma_{a}^{\dagger}=\gamma_{a},$ $a=1,\cdots4,$ $\gamma
_{4}=i\gamma_{0}$ and $\gamma_{5}=\gamma_{1}\gamma_{2}\gamma_{3}\gamma_{4}.$}.
The field strength
\begin{align*}
R  &  =d\omega+\omega^{2}\\
&  =\frac{1}{2}dx^{\mu}\wedge dx^{\nu}\biggl(\partial_{\mu}\omega_{\nu
}-\partial_{\nu}\omega_{\mu}+\left[  \omega_{\mu},\omega_{\nu}\right]
\biggr)\\
&  =\frac{1}{8}dx^{\mu}\wedge dx^{\nu}R_{\mu\nu ab}\gamma^{ab},
\end{align*}
transforms covariantly under the $SL(2,\mathbb{C})$ gauge transformation
$\Omega:$%
\begin{align*}
\omega &  \rightarrow\Omega~d~\Omega^{-1}+\Omega~\omega~\Omega^{-1},\\
R  &  \rightarrow\Omega~R~\Omega^{-1},
\end{align*}
where $\Omega=\exp\left(  \frac{1}{4}\Lambda_{ab}\gamma^{ab}\right)  $ and
$\Lambda_{ab}$ are the infinitesimal gauge parameters. Next, the vierbein $e$
defined by%
\[
e=dx^{\mu}e_{\mu}=dx^{\mu}e_{\mu}^{a}\gamma_{a},
\]
is introduced, which transforms under the $SL(2,\mathbb{C})$ according to
\[
e\rightarrow\Omega~e~\Omega^{-1}.
\]
The $SL(2,\mathbb{C})$ invariant gravitational action is then given by%
\begin{align*}
I  &  =\frac{1}{8}\int\limits_{M}Tr\left(  \left(  \alpha+\beta\gamma
_{5}\right)  e\wedge e\wedge R+\frac{\delta}{48}\gamma_{5}e\wedge e\wedge
e\wedge e\right) \\
&  =\frac{1}{16}\int\limits_{M}d^{4}x\,~\epsilon^{\mu\nu\kappa\lambda}\left(
\epsilon_{abcd}\left(  \beta e_{\mu}^{a}e_{\nu}^{b}R_{\kappa\lambda}%
^{\ ~cd}+\frac{\delta}{6}e_{\mu}^{a}e_{\nu}^{b}e_{\kappa}^{c}e_{\lambda}%
^{d}\right)  -2\alpha e_{\mu}^{a}e_{\nu}^{b}R_{\kappa\lambda}^{\ ~ab}\right)
.
\end{align*}
After the $\omega_{\mu}^{~ab}$ field is integrated out, this gives
\[
I=\frac{1}{4}%
{\displaystyle\int\limits_{M}}
d^{4}x~e\left(  R+\delta\right)  ,
\]
which is the Einstein-Hilbert action plus a cosmological constant. Notice that
the term with the coefficient $\alpha$ is of the form $\epsilon^{\mu\nu
\kappa\lambda}R_{\mu\nu\kappa\lambda\text{ }}$ and will vanish on shell by the
symmetries of the Riemann tensor. The invariance of the action under
$SL(2,\mathbb{C})$ transformations can be easily verified as $\Omega$ commutes
with $\gamma_{5}.$

In noncommutative geometry where the star product replaces ordinary products,
the group $SL(2,\mathbb{C})$ is first extended to $GL(2,\mathbb{C}) $ so that
the star product of two group elements is a group element. The field $e$ is
not preserved under group transformations
\[
e\rightarrow\Omega\ast e\ast\Omega_{\ast}^{-1},
\]
where $\Omega\ast\Omega_{\ast}^{-1}=1$. It is easily verified that the field
$e$ will become complex and must be replaced with the field $L$ defined by%
\[
L=dx^{\mu}\left(  e_{\mu}^{a}+i\gamma_{5}f_{\mu}^{a}\right)  \gamma_{a},
\]
which transforms properly under $L\rightarrow\Omega\ast L\ast\Omega_{\ast
}^{-1}.$ It is therefore necessary before studying any noncommutative
generalization to determine whether the gravitational theory with the field
$e$ replaced by $L$ is well defined. At first this idea does not seem to be
very promising because there will be two vierbeins $e_{\mu}^{a}$ and $f_{\mu
}^{a}$ and only one spin-connection $\omega_{\mu}^{~ab}.$ In the
Einstein-Hilbert action given above, the field $\omega_{\mu}^{~ab}$ appears
quadratically and can be determined exactly from its equation of motion as
function of the vierbein, and this is equivalent to performing a Gaussian
integration. The question we have to address is whether the couplings of
$\omega_{\mu}^{~ab}$ to $e_{\mu}^{a}$ and $f_{\mu}^{a}$ will be in such a way
as to insure the dynamical propagation of both fields. What is needed is to
get correct couplings for two symmetric tensors and two antisymmetric tensors
that could be formed out of $e_{\mu}^{a}$ and $f_{\mu}^{a}$. One combination
of the antisymmetric tensors could be gauged away by the $SL(2,\mathbb{C})$
invariance of the action. Moreover, because of the diffeomorphism invariance
of the full action, one combination of the symmetric tensors would correspond
to the massless graviton. The other symmetric combination would then
correspond to a massive graviton coupled to a scalar field (dilaton like). The
remaining antisymmetric field will be massive. In other words, this complex
gravity should be equivalent to bigravity \cite{strong}, \cite{damour},
\cite{georgi}, \cite{spontan} and yield the interaction of a massless graviton
coupled to a massive graviton and to a scalar field and a massive
antisymmetric tensor. It is essential to have $\omega_{\mu}^{~ab}$ generate
the correct kinetic energies for the two tensors. Happily, we shall show that
this is indeed the case, and remarkably there exists a coupling of the complex
vierbein $L$ to the curvature tensor that gives precisely the desired form
with correct signs. As mentioned before, the metric tensor is not introduced
apriori but results as a combination of the two fields $e_{\mu}^{a}$ and
$f_{\mu}^{a}.$ To deform the action so that ordinary products are replaced
with star products it is necessary to extend the group $SL(2,\mathbb{C})$ to
$GL(2,\mathbb{C}). $ Chiral rotations are present in $GL(2,\mathbb{C})$ and
this further rstricts the form of the invariant action. The invariant action
taken in the commutative case has to be modified. In this case it will be
necessary to impose a torsion free constraint on the complex vierbein $L$. The
$GL(2,\mathbb{C})$ gauge fields have to be determined by solving the torsion
free constraint instead of solving the equations of motion. Again this can
only be done perturbatively but it is relatively easy to to evaluate the
deformed action. It is also possible to derive the Seiberg-Witten map
\cite{SW}, \cite{wess} between the deformed and undeformed gauge fields and
complex vierbein. In contrast to earlier approaches we shall show that the
deformed action could be obtained without the use of this map and that its
form is manageable.

The plan of this paper is as follows. In section 2 we propose the action for
complex gravity in terms of the field $L.$ In section 3 we eliminate the field
$\omega_{\mu}^{~ab}$ in terms of $e_{\mu}^{a}$ and $f_{\mu}^{a}$ and show that
both tensors obtain the correct kinetic and mass terms. In section 4 we extend
the complex gravitational action to the noncommutative case where ordinary
products are replaced with star products. We also give transformations of the
deformed fields, the deformed action as well as the Seiberg-Witten map.
Section 5 contains the conclusion and some comments.

\section{ Gravity with a complex vierbein}

We start by considering the $SL(2,\mathbb{C})$ gauge connection $\omega$ and
the field $L$ transforming under $SL(2,\mathbb{C})$ as $L\rightarrow
\Omega~L~\Omega^{-1}.$ A generalization of the Einstein-Hilbert action for the
complex field
\[
L=dx^{\mu}\left(  e_{\mu}^{a}+if_{\mu}^{a}\gamma_{5}\right)  \gamma_{a},
\]
is given by
\[
I_{1}=\frac{1}{8}\int\limits_{M}Tr\biggl(\left(  \alpha+\beta\gamma
_{5}\right)  \left(  L+iL^{\prime}\right)  \wedge\left(  L-iL^{\prime}\right)
\wedge R\biggr),
\]
where
\begin{align*}
L^{\prime}  &  =dx^{\mu}\left(  e_{\mu}^{a}-if_{\mu}^{a}\gamma_{5}\right)
\gamma_{a}\\
&  =-CL^{T}C^{-1}%
\end{align*}
with $C$ being the charge conjugation matrix with the property $C\gamma
_{a}^{T}C^{-1}=-\gamma_{a}.$ Under $SL(2,\mathbb{C})$ gauge transformations
the field $L^{\prime}$ transforms as $L^{\prime}\rightarrow\Omega L^{\prime
}\Omega^{-1}.$ Notice that this action is Hermitian because
\begin{align*}
\omega_{\mu}^{\dagger}  &  =-\omega_{\mu},\;\;\;\;\;\;R_{\kappa\lambda
}^{\dagger}=-R_{\kappa\lambda},\\
\;L_{\mu}^{\dagger}  &  =L_{\mu},\;\;\;\;L_{\mu}^{\dagger\prime}=L_{\mu
},\;\;\;\;\;\gamma_{5}L_{\mu}=-L_{\mu}\gamma_{5}.
\end{align*}
It is possible to construct a different action where the combination $\left(
L_{\mu}L_{\nu}^{\prime}+L_{\mu}^{\prime}L_{\nu}\right)  $ replaces $\left(
L_{\mu}L_{\nu}+L_{\mu}^{\prime}L_{\nu}^{\prime}\right)  $. This would yield
the tensor combination $\left(  e_{\mu}^{a}e_{\nu}^{b}-f_{\mu}^{a}f_{\nu}%
^{b}\right)  $ instead of $\left(  e_{\mu}^{a}e_{\nu}^{b}+f_{\mu}^{a}f_{\nu
}^{b}\right)  $ which is undesirable result as it gives the wrong sign for the
kinetic energy of the massive graviton in the action. There are many
possibilities for the cosmological constant and mass terms. We shall choose a
combination of terms such that it would be possible to set the cosmological
constant to zero, have the linear terms in the fields $e_{\mu}^{a}$ and
$f_{\mu}^{a}$ vanish, and to get the Fierz-Pauli form \cite{FP} for the mass
of the spin-2 field. This is given by
\begin{align*}
&  I_{2}=\frac{1}{192}\int\limits_{M}Tr\biggl(\alpha_{1}\gamma_{5}\left(
L\wedge L\wedge L\wedge L+L\wedge L^{\prime}\wedge L\wedge L^{\prime}\right)
\biggr.\\
&  \qquad\biggl.+\frac{i}{8}\left(  L\wedge L^{\prime}-L^{\prime}\wedge
L\right)  \wedge\left(  \alpha_{2}\left(  L+L^{\prime}\right)  \wedge\left(
L+L^{\prime}\right)  -\alpha_{3}\left(  L-L^{\prime}\right)  \wedge\left(
L-L^{\prime}\right)  \right)  \biggr).
\end{align*}
To evaluate this action, we first expand it in terms of the component fields
$e_{\mu}^{a},$ $f_{\mu}^{a}$ and $\omega_{\mu}^{~ab}$ and then simplify the
Clifford algebra. The full action $I=I_{1}+I_{2}$ simplifies to%
\begin{align*}
I=  &  \frac{1}{2}\int\limits_{M}d^{4}x\;\epsilon^{\mu\nu\kappa\lambda
}\biggl(\epsilon_{abcd}\left(  \beta\left(  e_{\mu}^{a}e_{\nu}^{b}+f_{\mu}%
^{a}f_{\nu}^{b}\right)  +2\alpha e_{\mu}^{a}f_{\nu}^{b}\right)  R_{\kappa
\lambda}^{\quad cd}\biggr.\\
&  \qquad\qquad\qquad\left.  -2\bigl(\alpha\left(  e_{\mu}^{a}e_{\nu}%
^{b}+f_{\mu}^{a}f_{\nu}^{b}\right)  +2\beta e_{\mu}^{a}f_{\nu}^{b}%
\bigr)R_{\kappa\lambda}^{\quad ab}\right. \\
&  \qquad\qquad\qquad+\frac{1}{4!}\epsilon_{abcd}\alpha_{1}\left(  e_{\mu}%
^{a}e_{\nu}^{b}e_{\kappa}^{c}e_{\lambda}^{d}+f_{\mu}^{a}f_{\nu}^{b}f_{\kappa
}^{c}f_{\lambda}^{d}\right) \\
&  \qquad\qquad\qquad\biggl.+\frac{1}{4!}\epsilon_{abcd}\left(  \alpha
_{2}e_{\mu}^{a}e_{\nu}^{b}e_{\kappa}^{c}f_{\lambda}^{d}+\alpha_{3}f_{\mu}%
^{a}f_{\nu}^{b}f_{\kappa}^{c}e_{\lambda}^{d}\right)  \biggr).
\end{align*}
The field $\omega_{\mu}^{~ab}$ appears quadratically. This means that it can
be eliminated from the action by a Gaussian integration. Alternatively, we can
solve the $\omega_{\mu}^{~ab}$ equations of motion and substitute the value of
$\omega_{\mu}^{~ab}$ back into the action. In general this would require
inverting the tensor operator
\[
\epsilon^{\mu\nu\kappa\lambda}\biggl(\epsilon_{abcd}\bigl(\beta\left(  e_{\mu
}^{a}e_{\nu}^{b}+f_{\mu}^{a}f_{\nu}^{b}\right)  +2\alpha e_{\mu}^{a}f_{\nu
}^{b}\bigr)-2\bigl(\alpha\left(  e_{\mu}^{c}e_{\nu}^{d}+f_{\mu}^{c}f_{\nu}%
^{d}\right)  +2\beta e_{\mu}^{c}f_{\nu}^{d}\bigr)\biggr)
\]
This step could only be done perturbatively as function of $e_{a}^{\mu}$ and
$f_{a}^{\mu}$, the inverses of $e_{\mu}^{a}$ and $f_{\mu}^{a}.$ In fact the
analysis is fairly complicated, and in order to determine the dynamical
degrees of freedom of the system, it is essential to study the linearized
approximation. This is done by expanding $e_{\mu}^{a}$ and $f_{\mu}^{a}$
around a flat background by writing \cite{strong}
\begin{align*}
e_{\mu}^{a}  &  =c_{1}\delta_{\mu}^{a}+\overline{e}_{\mu}^{a}\\
f_{\mu}^{a}  &  =c_{2}\delta_{\mu}^{a}+\overline{f}_{\mu}^{a}%
\end{align*}
where $c_{1}$ and $c_{2}$ are parameters. Keeping only up to the bilinear
terms in $\overline{e}_{\mu}^{a}$ and $\overline{f}_{\mu}^{a}$ we obtain%
\begin{align*}
I  &  =2\int d^{4}x\biggl(-\left(  \beta\left(  c_{1}^{2}+c_{2}^{2}\right)
+2\alpha c_{1}c_{2}\right)  \left(  \omega_{dce}\omega_{ced}+\omega_{e}%
\omega_{e}\right)  \biggr.\\
&  \qquad\quad+\delta_{bcd}^{\nu\kappa\lambda}\partial_{\kappa}\omega_{\lambda
cd}\left(  \left(  \beta c_{1}+\alpha c_{2}\right)  \overline{e}_{\nu}%
^{b}+\left(  \alpha c_{1}+\beta c_{2}\right)  \overline{f}_{\nu}^{b}\right) \\
&  \qquad\quad-\left(  \alpha\left(  c_{1}^{2}+c_{2}^{2}\right)  +2\beta
c_{1}c_{2}\right)  \epsilon^{ab\kappa\lambda}\omega_{\kappa ae}\omega_{\lambda
eb}\\
&  \quad\qquad-2\epsilon^{a\nu\kappa\lambda}\partial_{\kappa}\omega_{\lambda
ab}\left(  \left(  \alpha c_{1}+\beta c_{2}\right)  \overline{e}_{\nu}%
^{b}+\left(  \beta c_{1}+\alpha c_{2}\right)  \overline{f}_{\nu}^{b}\right) \\
&  \qquad\quad+\left(  \alpha_{1}\left(  c_{1}^{4}+c_{2}^{4}\right)
+\alpha_{2}c_{1}^{3}c_{2}+\alpha_{3}c_{1}c_{2}^{3}\right) \\
&  \quad\qquad+\left(  4\alpha_{1}c_{1}^{3}+3\alpha_{2}c_{1}^{2}c_{2}%
+\alpha_{3}c_{2}^{3}\right)  \overline{e}+\left(  4\alpha_{1}c_{2}^{3}%
+3\alpha_{3}c_{1}c_{2}^{2}+\alpha_{2}c_{1}^{3}\right)  \overline{f}\\
&  \qquad\quad+3\delta_{\mu\nu}^{ab}\left(  \left(  2\alpha_{1}c_{1}%
^{2}+\alpha_{2}c_{1}c_{2}\right)  \overline{e}_{\mu}^{a}\overline{e}_{\nu}%
^{b}+\left(  2\alpha_{1}c_{2}^{2}+\alpha_{3}c_{1}c_{2}\right)  \overline
{f}_{\mu}^{a}\overline{f}_{\nu}^{b}\right. \\
&  \qquad\qquad\qquad\quad\biggl.\left.  +\left(  \alpha_{2}c_{1}^{2}%
+\alpha_{3}c_{2}^{2}\right)  \overline{e}_{\mu}^{a}\overline{f}_{\nu}%
^{b}\right)  +\cdots\biggr)
\end{align*}
As a first step, we write the $\omega_{\mu}^{~ab}$ equation of motion, which
takes the form%
\begin{align*}
&  \biggl(\bigl(\beta\left(  c_{1}^{2}+c_{2}^{2}\right)  +2\alpha c_{1}%
c_{2}\bigr)\bigl(\omega_{d\lambda c}-\omega_{c\lambda d}+\delta_{c\lambda
}\omega_{d}-\delta_{d\lambda}\omega_{c}\bigr)\biggr.\\
&  \biggl.+\bigl(\alpha\left(  c_{1}^{2}+c_{2}^{2}\right)  +2\beta c_{1}%
c_{2}\bigr)\bigl(\epsilon^{ad\kappa\lambda}\omega_{\kappa ac}-\epsilon
^{ac\kappa\lambda}\omega_{\kappa ad}\bigr)\biggr)\\
&  =-\delta_{bcd}^{\nu\kappa\lambda}\partial_{\kappa}\biggl(\left(  \beta
c_{1}+\alpha c_{2}\right)  \overline{e}_{\nu}^{b}+\left(  \alpha c_{1}+\beta
c_{2}\right)  \overline{f}_{\nu}^{b}\biggr)\\
&  +\biggl(\epsilon^{c\nu\kappa\lambda}\partial_{\kappa}\bigl(\left(  \alpha
c_{1}+\beta c_{2}\right)  \overline{e}_{\nu}^{d}+\left(  \beta c_{1}+\alpha
c_{2}\right)  \overline{f}_{\nu}^{d}\bigr)-c\leftrightarrow d\biggr).
\end{align*}
This is a difficult equation to solve. To simplify the problem we first define
the tensor
\begin{align*}
X_{mab}^{npq}  &  =\frac{a}{2}\left(  \delta_{m}^{n}\delta_{ab}^{pq}%
+\delta^{np}\delta_{ab}^{qm}-\delta^{nq}\delta_{ab}^{pm}\right) \\
&  +\frac{b}{2}\left(  \epsilon_{abnp}\delta_{qm}-\epsilon_{abnq}\delta
_{pm}\right)  ,
\end{align*}
where
\begin{align*}
a  &  =\beta\left(  c_{1}^{2}+c_{2}^{2}\right)  +2\alpha c_{1}c_{2},\\
b  &  =\alpha\left(  c_{1}^{2}+c_{2}^{2}\right)  +2\beta c_{1}c_{2}.
\end{align*}
We then define the tensor
\[
Y_{mab}=X_{mab}^{npq}\omega_{npq},
\]
so that the $\omega_{mab}$ equation simplifies to
\[
Y_{d\lambda c}-Y_{c\lambda d}=-\partial_{\kappa}\left(  \delta_{bcd}%
^{\nu\kappa\lambda}E_{\nu b}-\epsilon_{c\nu\kappa\lambda}F_{\nu d}%
+\epsilon_{d\nu\kappa\lambda}F_{\nu c}\right)  ,
\]
where
\begin{align*}
E_{\nu}^{b}  &  =\left(  \beta c_{1}+\alpha c_{2}\right)  \overline{e}_{\nu
}^{b}+\left(  \alpha c_{1}+\beta c_{2}\right)  \overline{f}_{\nu}^{b},\\
F_{\nu}^{b}  &  =\left(  \alpha c_{1}+\beta c_{2}\right)  \overline{e}_{\nu
}^{b}+\left(  \beta c_{1}+\alpha c_{2}\right)  \overline{f}_{\nu}^{b}.
\end{align*}
We can easily solve for $Y_{cd\lambda}$ by a cyclic permutation of the $Y$
equation to obtain%
\begin{align*}
Y_{cd\lambda}  &  =\frac{1}{2}\biggl(\partial_{c}\left(  E_{d\lambda
}-E_{\lambda d}\right)  -\partial_{d}\left(  E_{c\lambda}+E_{\lambda
c}\right)  +\partial_{\lambda}\left(  E_{cd}+E_{dc}\right)  \biggr)\\
&  -\delta_{c\lambda}\left(  \partial_{b}E_{db}-\partial_{d}E\right)
+\delta_{d\lambda}\left(  \partial_{b}E_{cb}-\partial_{c}E\right)
+\epsilon_{d\lambda\kappa\nu}\partial_{\kappa}F_{\nu c},
\end{align*}
where $E=E_{bb}.$ We now define the inverse of the tensor $X_{mab}^{npq}$ by%
\[
\left(  X^{-1}\right)  _{rst}^{mab}X_{mab}^{npq}=\frac{1}{2}\delta_{r}%
^{n}\delta_{st}^{pq}.
\]
To find the inverse we write the most general rank 6 tensor antisymmetric in
$s$ and $t$ and in $p$ and $q$ \ then determine the coefficients from the
above constraint. After a lengthy calculation we obtain
\begin{align*}
\left(  X^{-1}\right)  _{rst}^{mab}  &  =\frac{1}{2\left(  a^{2}-b^{2}\right)
}\biggl(a\bigl(\delta_{r}^{m}\delta_{st}^{ab}-\frac{1}{2}\delta_{rs}%
\delta_{mt}^{ab}+\frac{1}{2}\delta_{rt}\delta_{ms}^{ab}\bigr)\biggr.\\
&  -b\bigl(\epsilon_{stma}\delta_{br}-\epsilon_{stmb}\delta_{ar}%
+\epsilon_{mabs}\delta_{tr}-\epsilon_{mabt}\delta_{sr}\bigr.\\
&  \qquad\qquad\biggr.\bigl.-\frac{1}{2}\epsilon_{rsta}\delta_{mb}+\frac{1}%
{2}\epsilon_{rstb}\delta_{ma}\bigr)\biggr).
\end{align*}
We can then write
\[
\omega_{rst}=\left(  X^{-1}\right)  _{rst}^{mab}Y_{mab},
\]
and after some algebra one finds
\begin{align*}
\omega_{rst}  &  =\frac{1}{2\left(  a^{2}-b^{2}\right)  }\biggl(\partial
_{r}\left(  aE_{st}-bF_{st}\right)  -\partial_{s}\left(  aE_{rt}%
-bF_{rt}\right)  \biggr.\\
&  \qquad\qquad\qquad\biggl.+\epsilon_{st\mu\nu}\partial_{\mu}\left(  aF_{\nu
r}-bE_{\nu r}\right)  +\delta_{rs}\epsilon_{t\mu\nu m}\partial_{\mu}\left(
aF_{\nu m}-bE_{\nu m}\right)  -s\leftrightarrow t\biggr).
\end{align*}
This expression simplifies by noting that
\begin{align*}
aE_{st}-bF_{st}  &  =\left(  \beta^{2}-\alpha^{2}\right)  \left(  c_{1}%
^{2}-c_{2}^{2}\right)  g_{\nu a},\\
bE_{rt}-aF_{rt}  &  =\left(  \beta^{2}-\alpha^{2}\right)  \left(  c_{1}%
^{2}-c_{2}^{2}\right)  h_{\nu a},
\end{align*}
where we have defined
\begin{align*}
g_{\nu a}  &  =\left(  c_{1}\overline{e}_{\nu a}-c_{2}\overline{f}_{\nu
a}\right)  ,\\
h_{\nu a}  &  =\left(  -c_{2}\overline{e}_{\nu a}+c_{1}\overline{f}_{\nu
a}\right)  .
\end{align*}
We finally have
\begin{align*}
\omega_{rst}  &  =\frac{1}{2\left(  c_{1}^{2}-c_{2}^{2}\right)  }%
\biggl(\partial_{r}g_{st}-\partial_{s}\left(  g_{rt}+g_{tr}\right)
+\epsilon_{st\mu\nu}\partial_{\mu}h_{\nu r}+\delta_{rs}\epsilon_{t\mu\nu
m}\partial_{\mu}h_{\nu m}-s\leftrightarrow t\biggr),\\
\omega_{t}  &  =\frac{1}{\left(  c_{1}^{2}-c_{2}^{2}\right)  }\left(
-\partial_{r}g_{tr}+\partial_{t}g+\frac{1}{2}\epsilon_{t\mu\nu r}\partial
_{\mu}h_{\nu r}\right)  ,
\end{align*}
where $\omega_{t}=\omega_{rrt}.$ To avoid degeneracy we shall impose the
following constraints on the parameters $\alpha,~\beta,~c_{1}$ and $c_{2}:$%
\[
\alpha\neq\beta,\;c_{1}\neq c_{2}.
\]
Substituting these expressions back into the action, we find that the
antisymmetric part of $g_{ab}$ decouples, while both the symmetric and
antisymmetric parts of $h_{ab}$ couple and acquire kinetic energies. We
therefore write
\begin{align*}
g_{\mu\nu}  &  =\frac{1}{2}\left(  s_{\mu\nu}+a_{\mu\nu}\right)  ,\\
h_{\mu\nu}  &  =\left(  l_{\mu\nu}+B_{\mu\nu}\right)  ,
\end{align*}
where $s_{\mu\nu}$, $l_{\mu\nu}$ and $a_{\mu\nu}$, $B_{\mu\nu}$ are
respectively the symmetric and antisymmetric parts of $g_{\mu\nu}$ and
$h_{\mu\nu}.$ Keeping only up to bilinear terms, the action reduces to%
\begin{align*}
I  &  =-\frac{4}{\left(  c_{1}^{2}-c_{2}^{2}\right)  ^{2}}\int d^{4}%
x~\biggl(\partial_{\mu}s_{\nu\kappa}\partial_{\mu}s_{\nu\kappa}-2\partial
_{\mu}s_{\mu\kappa}\partial_{\nu}s_{\nu\kappa}+2\partial_{\mu}s_{\mu\nu
}\partial_{\nu}s-\partial_{\mu}s\partial_{\mu}s\biggr.\\
&  \qquad\qquad\qquad\qquad+\partial_{\mu}l_{\nu\kappa}\partial_{\mu}%
l_{\nu\kappa}-2\partial_{\mu}l_{\mu\kappa}\partial_{\nu}l_{\nu\kappa
}+2\partial_{\mu}l_{\mu\nu}\partial_{\nu}l-\partial_{\mu}l\partial_{\mu}l\\
&  \qquad\qquad\qquad\qquad\biggl.+\partial_{\mu}B_{\nu\kappa}\partial_{\mu
}B_{\nu\kappa}-2\partial_{\mu}B_{\mu\kappa}\partial_{\nu}B_{\nu\kappa
}\biggr)\\
&  +\int d^{4}x~\biggl(\alpha_{1}\left(  c_{1}^{4}+c_{2}^{4}\right)
+\alpha_{2}\left(  c_{1}^{3}c_{2}\right)  +\alpha_{3}\left(  c_{1}c_{2}%
^{3}\right)  \biggr)\\
&  +\frac{1}{\left(  c_{1}^{2}-c_{2}^{2}\right)  }\int d^{4}x~\biggl(4\left(
\alpha_{1}\left(  c_{1}^{4}+c_{2}^{4}\right)  +\alpha_{2}c_{1}^{3}c_{2}%
+\alpha_{3}c_{1}c_{2}^{3}\right)  g\biggr.\\
&  \hspace{1in}\quad+\biggl.\left(  4\alpha_{1}c_{1}c_{2}\left(  c_{1}%
^{2}+c_{2}^{2}\right)  +\alpha_{2}c_{1}^{2}\left(  c_{1}^{2}+3c_{2}%
^{2}\right)  +\alpha_{3}c_{2}^{2}\left(  c_{2}^{2}+3c_{1}^{2}\right)  \right)
h\biggr)\\
&  +\frac{1}{\left(  c_{1}^{2}-c_{2}^{2}\right)  ^{2}}\int d^{4}x~\delta
_{\mu\nu}^{\kappa\lambda}\biggl(2\bigl(\alpha_{1}\left(  c_{1}^{4}+c_{2}%
^{4}\right)  +\alpha_{2}\left(  c_{1}^{3}c_{2}\right)  +\alpha_{3}\left(
c_{1}c_{2}^{3}\right)  \bigr)g_{\mu\kappa}g_{\nu\lambda}\biggr.\\
&  \hspace{1.2in}+\bigl(\left(  4\alpha_{1}c_{1}c_{2}\left(  c_{1}^{2}%
+c_{2}^{2}\right)  +\alpha_{2}c_{1}^{2}\left(  c_{1}^{2}+3c_{2}^{2}\right)
+\alpha_{3}c_{2}^{2}\left(  c_{2}^{2}+3c_{1}^{2}\right)  \right)  g_{\mu
\kappa}h_{\nu\lambda}\bigr)\\
&  \hspace{1.2in}\quad\quad+\biggl.\left(  4\alpha_{1}c_{1}^{2}c_{2}%
^{2}+\left(  \alpha_{2}+\alpha_{3}\right)  c_{1}c_{2}\left(  c_{1}^{2}%
+c_{2}^{2}\right)  \right)  h_{\mu\kappa}h_{\nu\lambda}\biggr),
\end{align*}
where $g=g_{\mu\mu},$ $h=h_{\mu\mu}.$ By setting the cosmological term and
linear terms in $g$ and $h$ to zero we get three equations in the three
parameters $\alpha_{1},$ $\alpha_{2}$ and $\alpha_{3}$. Only two of the
equations are independent, and they are%
\begin{align*}
\alpha_{1}\left(  c_{1}^{4}+c_{2}^{4}\right)  +\alpha_{2}c_{1}^{3}c_{2}%
+\alpha_{3}c_{1}c_{2}^{3}  &  =0,\\
4\alpha_{1}c_{1}^{3}+3\alpha_{2}c_{1}^{2}c_{2}+\alpha_{3}c_{2}^{3}  &  =0,
\end{align*}
These can be easily solved to determine $\alpha_{2}$ and $\alpha_{3}$ in terms
of $\alpha_{1}:$%
\begin{align*}
\alpha_{2}  &  =\frac{k^{4}-3}{2k}\alpha_{1},\\
\alpha_{3}  &  =\frac{1-3k^{4}}{2k^{3}}\alpha_{1},
\end{align*}
where
\[
k=\frac{c_{2}}{c_{1}}\neq1.
\]
With this solution one immediately finds that both the mass term $\delta
_{\mu\nu}^{\kappa\lambda}g_{\mu\kappa}g_{\nu\lambda}$ and the mixing term
$\delta_{\mu\nu}^{\kappa\lambda}g_{\mu\kappa}h_{\nu\lambda}$ vanish. There is
however a mass term for $h_{\mu\nu}$
\[
\frac{3\alpha_{1}\left(  k^{4}+1\right)  }{2k^{2}}\delta_{\mu\nu}%
^{\kappa\lambda}h_{\mu\kappa}h_{\nu\lambda},
\]
which is of the Fierz-Pauli type \cite{FP}. The order of the mass term can be
tuned by adjusting the parameters $\alpha_{1}$ and $k.$ Note that both the
symmetric field $l_{\mu\nu}$ and the antisymmetric field $B_{\mu\nu}$ acquire
mass. It is not unexpected that the graviton field remains massless as this is
protected by diffeomorphism invariance. However, it is remarkable that through
the coupling of the spin-connection $\omega_{\mu ab}$ the correct kinetic
energies for both fields $g_{\mu\nu}$ and $h_{\mu\nu}$ are generated. The
degrees of freedom of this system are well defined. The graviton, corresponds
to a massless spin-2 field has two dynamical degrees of freedom, while the
field $h_{\mu\nu}$ corresponds to a massive spin-2 coupled to a dilaton and
has 6 degrees of freedom \cite{damour}. The dilaton coupling can only be seen
by going to higher order terms as it couples to curvature terms \cite{georgi},
\cite{spontan}. To have a closed form for the fully non-linear theory, it
would be necessary to define an inverse for the tensor $\left(  e_{\mu}%
^{a}e_{\nu}^{b}+f_{\mu}^{a}f_{\nu}^{b}-e_{\nu}^{a}e_{\mu}^{b}-f_{\nu}%
^{a}f_{\mu}^{b}\right)  $ \ so as to express the action in terms of this inverse.

Much work remains to be done to fully understand this theory and to determine
its full coupling at the non-linear level, but the above results are very
encouraging and strongly indicate that this theory is consistent. It is also
very geometrical based, on the gauge principle where all terms in the action
are four-forms thus avoiding the use of a density factor to guarantee
invariance under general coordinate transformations. It would be very
interesting to find some particular solutions to the full field equations such
as generalizations of the Schwarzschild or de Sitter solutions.

\section{Noncommutative Deformed Gravity}

The construction of the complex gravity action proposed in the last section
suggests that it could be easily generalized to the noncommutative case where
the coordinates of space-time do not commute%
\[
\left[  x^{\mu},x^{\nu}\right]  =i\theta^{\mu\nu}%
\]
where $\theta^{\mu\nu}$ are deformation parameters. An immediate step is to
extend the $SL(2,\mathbb{C)}$ group to $GL(2,\mathbb{C)}$. This is necessary
because the commutator in a star product involves both ordinary commutators
and anticommutators as can be seen from the relation%
\[
A\ast B-B\ast A=\left[  A,B\right]  _{\left(  \ast,\text{even}\right)
}+\left\{  A,B\right\}  _{\left(  \ast,\text{odd}\right)  },
\]
where%
\begin{align*}
\left[  A,B\right]  _{\left(  \ast,\text{even}\right)  } &  =\left[
A,B\right]  +\left(  \frac{i}{2}\right)  ^{2}\theta^{\mu\nu}\theta
^{\kappa\lambda}\left[  \partial_{\mu}\partial_{\kappa}A,\partial_{\nu
}\partial_{\lambda}B\right]  +O(\theta^{4}),\\
\left\{  A,B\right\}  _{\left(  \ast,\text{odd}\right)  } &  =\frac{i}%
{2}\theta^{\mu\nu}\left\{  \partial_{\mu}A,\partial_{\nu}B\right\}  +\left(
\frac{i}{2}\right)  ^{3}\theta^{\mu\nu}\theta^{\kappa\lambda}\theta
^{\alpha\beta}\left\{  \partial_{\mu}\partial_{\kappa}\partial_{\alpha
}A,\partial_{\nu}\partial_{\lambda}\partial_{\beta}B\right\}  +O(\theta^{5}).
\end{align*}
With this modification we first define the $GL(2,\mathbb{C)}$ gauge field
$\widetilde{A}_{\mu}$%
\[
\widetilde{A}=dx^{\mu}\left(  i\left(  \widetilde{a}_{\mu}+\widetilde{b}_{\mu
}\gamma_{5}\right)  +\frac{1}{4}\widetilde{\omega}_{\mu ab}\gamma^{ab}\right)
,
\]
satisfying the condition $\widetilde{A}_{\mu}^{\dagger}=-\widetilde{A}_{\mu}$
and transforming under a gauge transformation according to
\[
\widetilde{A}\rightarrow\widetilde{\Omega}\ast\widetilde{A}\ast\widetilde
{\Omega}_{\ast}^{-1}+\widetilde{\Omega}\ast d~\widetilde{\Omega}_{\ast}^{-1}%
\]
where $\widetilde{\Omega}=e^{\widetilde{\lambda}}$ with
\[
\widetilde{\lambda}=i\left(  \widetilde{\alpha}+\widetilde{\beta}\gamma
_{5}\right)  +\frac{1}{4}\widetilde{\lambda}_{ab}\gamma^{ab}%
\]
One can easily verify that these transformation close as both the commutators
and anticommutators of $\gamma_{ab}$ with $\gamma_{c}$ and $\gamma_{c}%
\gamma_{5}$ are proportional to $\gamma_{d}$ and $\gamma_{d}\gamma_{5}$.The
field strength is%
\begin{align*}
\widetilde{F} &  =\frac{1}{2}dx^{\mu}\wedge dx^{\nu}\widetilde{F}_{\mu\nu},\\
\widetilde{F}_{\mu\nu} &  =\partial_{\mu}\widetilde{A}_{\nu}-\partial_{\nu
}\widetilde{A}_{\mu}+\widetilde{A}_{\mu}\ast\widetilde{A}_{\nu}-\widetilde
{A}_{\nu}\ast\widetilde{A}_{\mu},
\end{align*}
transforming according to
\[
\widetilde{F}_{\mu\nu}=\widetilde{\Omega}\ast\widetilde{F}_{\mu\nu}%
\ast\widetilde{\Omega}_{\ast}^{-1}%
\]
The field $\widetilde{L}_{\text{ }}$ is defined as before
\[
\widetilde{L}=dx^{\mu}\left(  \widetilde{e}_{\mu}^{a}+i\gamma_{5}\widetilde
{f}_{\mu}^{a}\right)  \gamma_{a},
\]
and transforms according to
\[
\widetilde{L}\rightarrow\widetilde{\Omega}\ast\widetilde{L}\ast\widetilde
{\Omega}_{\ast}^{-1}.
\]
Unlike the commutative case the field
\[
L^{\prime}=-CL^{T}C^{-1}%
\]
transforms as
\[
\widetilde{L^{\prime}}\rightarrow\widetilde{\Omega^{\prime}}\ast
\widetilde{L^{\prime}}\ast\widetilde{\Omega^{\prime}}_{\ast}^{-1}%
\]
where $\widetilde{\Omega^{\prime}}=e^{\widetilde{\lambda^{\prime}}}$ with
\[
\widetilde{\lambda^{\prime}}=-i\left(  \widetilde{\alpha}+\widetilde{\beta
}\gamma_{5}\right)  +\frac{1}{4}\widetilde{\lambda}_{ab}\gamma^{ab}.
\]
It is therefore not possible to construct a group invariant using both
$\widetilde{L}$ and $\widetilde{L^{\prime}}$ as for $SL(2,\mathbb{C)}$ where
$\lambda$ and $\lambda^{\prime}$ coincide. Therefore we are forced to use only
the fields $\widetilde{L}$ and $\widetilde{A}$ to construct an action
invariant under $GL(2,\mathbb{C)}$. It can be easily seen from the analysis
given in the last section that since the field $\widetilde{L^{\prime}}$ can
not used, the coupling of $\ \widetilde{\omega}_{\mu ab}$ to $\widetilde
{e}_{\mu}^{a}$ and $\widetilde{f}_{\mu}^{a}$ insure only the propagation of
one combination of $\widetilde{e}_{\mu}^{a}$ and $\widetilde{f}_{\mu}^{a}.$
\ It is immediate to write the deformed four dimensional gravitational action
invariant under the noncommutative $GL(2,\mathbb{C)}$ gauge transformations:%
\begin{align*}
&  \widetilde{I}=\int\limits_{M}d^{4}x~\epsilon^{\mu\nu\kappa\lambda
}Tr\biggl(\left(  \alpha_{1}+\beta_{1}\gamma_{5}\right)  \left(  \widetilde
{L}\ast\widetilde{L}\ast\widetilde{F}\right)  \biggr)\\
+ &  \int\limits_{M}d^{4}x\;\epsilon^{\mu\nu\kappa\lambda}Tr\biggl(\left(
\alpha_{2}+\beta_{2}\gamma_{5}\right)  \bigl(\widetilde{L}\ast\widetilde
{L}\ast\widetilde{L}\ast\widetilde{L}\bigr)\biggr)
\end{align*}
To this it is possible but not necessary to  add the torsion-free constraint
\[
\widetilde{T}=d\widetilde{L}+\widetilde{A}\ast\widetilde{L}+\widetilde{L}%
\ast\widetilde{A}=0
\]
which can be decomposed in terms of components and then solved.

We first determine the infinitesimal gauge transformations of the gauge
fields
\[
\delta\widetilde{A}=-d\widetilde{\lambda}+\widetilde{\lambda}\ast\widetilde
{A}-\widetilde{A}\ast\widetilde{\lambda},
\]
where $\widetilde{\Omega}=e^{\widetilde{\lambda}}$ and $\widetilde{\lambda
}=i(\widetilde{\alpha}+\gamma_{5}\widetilde{\beta})+\frac{1}{4}\widetilde
{\lambda}^{ab}\gamma_{ab}.$ In terms of components this reads%
\begin{align*}
\delta\widetilde{a}_{\mu}  &  =-\partial_{\mu}\widetilde{\alpha}%
-\theta^{\kappa\lambda}\bigl(\partial_{\kappa}\widetilde{\alpha}%
\partial_{\lambda}\widetilde{a}_{\mu}+\partial_{\kappa}\widetilde{\beta
}\partial_{\lambda}\widetilde{b}_{\mu}+\frac{1}{8}\partial_{\kappa}%
\widetilde{\lambda}^{ab}\partial_{\lambda}\widetilde{\omega}_{\mu
ab}\bigr)+O(\theta^{3}),\\
\delta\widetilde{b}_{\mu}  &  =-\partial_{\mu}\widetilde{\beta}-\theta
^{\kappa\lambda}\bigl(\partial_{\kappa}\widetilde{\beta}\partial_{\lambda
}\widetilde{a}_{\mu}+\partial_{\kappa}\widetilde{\alpha}\partial_{\lambda
}\widetilde{b}_{\mu}-\frac{1}{16}\epsilon^{abcd}\partial_{\kappa}%
\widetilde{\lambda}_{ab}\partial_{\lambda}\widetilde{\omega}_{\mu
cd}\bigr)+O(\theta^{3}),\\
\delta\widetilde{\omega}_{\mu ab}  &  =-\left(  \partial_{\mu}\widetilde
{\lambda}_{ab}+\widetilde{\omega}_{\mu ac}\widetilde{\lambda}_{cb}%
-\widetilde{\omega}_{\mu bc}\widetilde{\lambda}_{ca}\right) \\
&  -\theta^{\kappa\lambda}\bigl(\left(  \partial_{\kappa}\widetilde{\alpha
}\partial_{\lambda}\widetilde{\omega}_{\mu ab}+\partial_{\kappa}%
\widetilde{\lambda}_{ab}\partial_{\lambda}\widetilde{a}_{\mu}\right)
+\frac{1}{2}\epsilon_{abcd}\left(  \partial_{\kappa}\widetilde{\beta}%
\partial_{\lambda}\widetilde{\omega}_{\mu cd}+\partial_{\kappa}\widetilde
{\lambda}_{cd}\partial_{\lambda}\widetilde{b}_{\mu}\right)  \bigr)\\
&  -\frac{1}{4}\theta^{\alpha\beta}\theta^{\gamma\delta}\bigl(\partial
_{\alpha}\partial_{\gamma}\widetilde{\omega}_{\mu ac}\partial_{\beta}%
\partial_{\delta}\widetilde{\lambda}_{cb}-\partial_{\alpha}\partial_{\gamma
}\widetilde{\omega}_{\mu bc}\partial_{\beta}\partial_{\delta}\widetilde
{\lambda}_{ca}\bigr)+O(\theta^{3}).
\end{align*}
Similarly the infinitesimal gauge transformation of the complex vierbein $L$
is given by
\[
\delta\widetilde{L}=\widetilde{\lambda}\ast\widetilde{L}-\widetilde{L}%
\ast\widetilde{\lambda},
\]
which in component form reads%
\begin{align*}
\delta\widetilde{e}_{\mu}^{a}  &  =\widetilde{\lambda}^{ac}\widetilde{e}_{\mu
}^{c}-\theta^{\gamma\delta}\bigl(\partial_{\gamma}\widetilde{\alpha}%
\partial_{\delta}\widetilde{e}_{\mu}^{a}-\frac{1}{8}\epsilon^{abcd}%
\partial_{\gamma}\widetilde{\lambda}_{bc}\partial_{\delta}\widetilde{f}_{\mu
}^{d}\bigr)\\
&  -\frac{1}{4}\theta^{\alpha\beta}\theta^{\gamma\delta}\bigl(\partial
_{\alpha}\partial_{\gamma}\widetilde{\lambda}^{ac}\partial_{\beta}%
\partial_{\delta}\widetilde{e}_{\mu c}\bigr)+O(\theta^{3}),\\
\delta\widetilde{f}_{\mu}^{a}  &  =\widetilde{\lambda}^{ac}\widetilde{f}_{\mu
}^{c}-\theta^{\gamma\delta}\bigl(\partial_{\gamma}\alpha\partial_{\delta
}\widetilde{f}_{\mu}^{a}+\frac{1}{8}\epsilon^{abcd}\partial_{\gamma}%
\widetilde{\lambda}_{bc}\partial_{\delta}\widetilde{e}_{\mu}^{d}\bigr)\\
&  -\frac{1}{4}\theta^{\alpha\beta}\theta^{\gamma\delta}\biggl(\partial
_{\alpha}\partial_{\gamma}\widetilde{\lambda}^{ac}\partial_{\beta}%
\partial_{\delta}\widetilde{f}_{\mu c}\biggr)+O(\theta^{3}).
\end{align*}
The components of the torsion constraints are%
\[
\widetilde{T}_{\mu\nu}=\widetilde{T}_{\mu\nu}^{a}\gamma_{a}+\widetilde{T}%
_{\mu\nu}^{a5}i\gamma_{5}\gamma_{a}=0
\]
where
\begin{align*}
\widetilde{T}_{\mu\nu}^{a}  &  =\left(  \partial_{\mu}\widetilde{e}_{\nu}%
^{a}+\frac{1}{2}\left\{  \widetilde{\omega}_{\mu ab},\;\widetilde{e}_{\nu}%
^{b}\right\}  _{\ast}-\frac{i}{4}\epsilon^{abcd}\left[  \widetilde{\omega
}_{\mu bc},\;\widetilde{f}_{\nu d}\right]  _{\ast}\right. \\
&  \left.  +i\left[  \widetilde{a}_{\mu},\;\widetilde{e}_{\nu}^{a}\right]
_{\ast}-\left\{  \widetilde{b}_{\mu},\;\widetilde{f}_{v}^{a}\right\}  _{\ast
}-\mu\leftrightarrow\nu\right) \\
\widetilde{T}_{\mu\nu}^{a}  &  =\left(  \partial_{\mu}\widetilde{f}_{\nu}%
^{a}+\frac{1}{2}\left\{  \widetilde{\omega}_{\mu ab},\;\widetilde{f}_{\nu}%
^{b}\right\}  _{\ast}+\frac{i}{4}\epsilon^{abcd}\left[  \widetilde{\omega
}_{\mu bc},\;\widetilde{e}_{\nu d}\right]  _{\ast}\right. \\
&  \left.  +i\left[  \widetilde{a}_{\mu},\;\widetilde{f}_{\nu}^{a}\right]
_{\ast}+\left\{  \widetilde{b}_{\mu},\;\widetilde{e}_{v}^{a}\right\}  _{\ast
}-\mu\leftrightarrow\nu\right)
\end{align*}
These equations simplify when written in terms of the complex field
\[
\widetilde{\mathbb{E}}_{\mu}^{a}=\widetilde{e}_{\mu}^{a}+i\widetilde{f}_{\mu
}^{a}%
\]
as they take the form%
\begin{align*}
0  &  =\left(  \partial_{\mu}\widetilde{\mathbb{E}}_{\nu}^{a}+\frac{1}%
{2}\left\{  \widetilde{\omega}_{\mu ab},\;\widetilde{\mathbb{E}}_{\nu}%
^{b}\right\}  _{\ast}-\frac{1}{4}\epsilon^{abcd}\left[  \widetilde{\omega
}_{\mu bc},\;\widetilde{\mathbb{E}}_{\nu d}\right]  _{\ast}\right. \\
&  \left.  +i\left[  \widetilde{a}_{\mu},\;\widetilde{\mathbb{E}}_{\nu}%
^{a}\right]  _{\ast}+i\left\{  \widetilde{b}_{\mu},\;\widetilde{\mathbb{E}%
}_{v}^{a}\right\}  _{\ast}-\mu\leftrightarrow\nu\right)
\end{align*}
as well as the complex conjugate equation.

We now determine the deformed action to second order in $\theta.$ The gauge
field strength is given by%
\[
\widetilde{F}_{\mu\nu}=i\left(  \widetilde{a}_{\mu\nu}+\gamma_{5}\widetilde
{b}_{\mu\nu}\right)  +\frac{1}{4}\widetilde{R}_{\mu\nu ab}\gamma^{ab}%
\]
where
\begin{align*}
\widetilde{a}_{\mu\nu}  &  =\partial_{\mu}\widetilde{a}_{\nu}-\partial_{\nu
}\widetilde{a}_{\mu}+i\left[  \widetilde{a}_{\mu},\;\widetilde{a}_{\nu
}\right]  _{\ast}+i\left[  \widetilde{b}_{\mu},\;\widetilde{b}_{\nu}\right]
_{\ast}+\frac{i}{8}\left[  \widetilde{\omega}_{\mu}^{\;ab},\;\widetilde
{\omega}_{vab}\right]  _{\ast}\\
\widetilde{b}_{\mu\nu}  &  =\partial_{\mu}\widetilde{b}_{\nu}-\partial_{\nu
}\widetilde{b}_{\mu}+i\left[  \widetilde{a}_{\mu},\;\widetilde{b}_{\nu
}\right]  _{\ast}+i\left[  \widetilde{b}_{\mu},\;\widetilde{a}_{\nu}\right]
_{\ast}-\frac{i}{8}\epsilon_{abcd}\left[  \widetilde{\omega}_{\mu}%
^{\;ab},\;\widetilde{\omega}_{v}^{\;cd}\right]  _{\ast}\\
\widetilde{R}_{\mu\nu ab}  &  =\partial_{\mu}\widetilde{\omega}_{vab}+i\left[
\widetilde{\omega}_{\mu ab},\;\widetilde{a}_{\nu}\right]  _{\ast}+\frac{i}%
{2}\epsilon_{abcd}\left[  \widetilde{b}_{\mu},\;\widetilde{\omega}_{v}%
^{\;cd}\right]  _{\ast}+\frac{1}{2}\left\{  \widetilde{\omega}_{\mu}%
^{\;ac},\;\widetilde{\omega}_{\nu c}^{\;\,b}\right\}  _{\ast}-\mu
\leftrightarrow\nu
\end{align*}
To determine the deformed action we first expand the combination%
\[
\widetilde{L}\ast\widetilde{L}=\frac{1}{2}dx^{\mu}\wedge dx^{\nu}%
\biggl(l_{\mu\nu ab}\gamma^{ab}+i\bigl(l_{\mu\nu}^{\left(  1\right)  }%
+\gamma_{5}l_{\mu\nu}^{\left(  5\right)  }\bigr)\biggr)
\]
where
\begin{align*}
l_{\mu\nu}^{\;ab}  &  =\left\{  \widetilde{e}_{\mu}^{a},\;\widetilde{e}%
_{v}^{b}\right\}  _{\ast}+\left\{  \widetilde{f}_{\mu}^{a},\;\widetilde
{f}_{\nu}^{b}\right\}  -\frac{i}{2}\epsilon^{abcd}\left(  \left[
\widetilde{e}_{\mu c},\;\widetilde{f}_{\nu d}\right]  _{\ast}-\mu
\leftrightarrow\nu\right) \\
l_{\mu\nu}^{\left(  1\right)  }  &  =-i\left[  \widetilde{e}_{\mu}%
^{a},\;\widetilde{e}_{va}\right]  _{\ast}-i\left[  \widetilde{f}_{\mu}%
^{a},\;\widetilde{f}_{va}\right]  _{\ast}\\
l_{\mu\nu}^{\left(  5\right)  }  &  =-\left\{  \widetilde{e}_{\mu}%
^{a},\;\widetilde{f}_{va}\;\right\}  _{\ast}+\left\{  \widetilde{e}_{\nu}%
^{a},\;\widetilde{f}_{\mu a}\;\right\}  _{\ast}.
\end{align*}
The kinetic part of the action then takes the form
\begin{align*}
\widetilde{I}_{1}  &  =%
{\displaystyle\int\limits_{M}}
d^{4}x~\epsilon^{\mu\nu\kappa\lambda}\biggl(-\alpha_{1}\bigl(l_{\mu\nu
}^{\left(  1\right)  }\ast\widetilde{a}_{\kappa\lambda}+l_{\mu\nu}^{\left(
5\right)  }\ast\widetilde{b}_{\kappa\lambda}+\frac{1}{2}l_{\mu\nu}^{\;ab}%
\ast\widetilde{R}_{\kappa\lambda ab}\bigr)\biggr.\\
&  \hspace{1in}\biggl.-\beta_{1}\bigl(l_{\mu\nu}^{\left(  1\right)  }%
\ast\widetilde{b}_{\kappa\lambda}+l_{\mu\nu}^{\left(  5\right)  }%
\ast\widetilde{a}_{\kappa\lambda}+\frac{1}{4}\epsilon_{abcd}\;l_{\mu\nu
}^{\;ab}\ast\widetilde{R}_{\kappa\lambda}^{\;\;cd}\bigr)\biggr).
\end{align*}
while the cosmological term gives%
\begin{align*}
\widetilde{I}_{2}  &  =%
{\displaystyle\int\limits_{M}}
d^{4}x~\epsilon^{\mu\nu\kappa\lambda}\biggl(-\alpha_{2}\bigl(l_{\mu\nu
}^{\left(  1\right)  }\ast l_{\kappa\lambda}^{\left(  1\right)  }+l_{\mu\nu
}^{\left(  5\right)  }\ast l_{\kappa\lambda}^{\left(  5\right)  }+2l_{\mu\nu
}^{\;ab}\ast l_{\kappa\lambda ab}\bigr)\biggr.\\
&  \biggl.-\beta_{2}\bigl(2l_{\mu\nu}^{\left(  1\right)  }\ast l_{\kappa
\lambda}^{\left(  5\right)  }-\epsilon_{abcd}\,\;l_{\mu\nu}^{\;ab}\ast
l_{\kappa\lambda}^{\;\;cd}\bigr)\biggr).
\end{align*}

\bigskip\ The Seiberg-Witten map \cite{SW}, \cite{wess} determining the
deformed gauge field in terms of the undeformed one is defined by
\[
\widetilde{A}\left(  g~A~g^{-1}+g~dg^{-1}\right)  =\widetilde{g}\ast
\widetilde{A}\left(  A\right)  \ast\widetilde{g}_{\ast}^{-1}+\widetilde{g}\ast
d\widetilde{g}_{\ast}^{-1}.
\]
Its solution is given by
\begin{align*}
\widetilde{A}_{\mu}  &  =A_{\mu}-\frac{i}{4}\theta^{\kappa\lambda}\left\{
A_{\kappa},\,\partial_{\lambda}A_{\mu}+F_{\lambda\mu}\right\}  +O(\theta
^{2}),\\
\widetilde{F}_{\mu\nu}  &  =F_{\mu\nu}+\frac{i}{4}\theta^{\kappa\lambda
}\left(  2\left\{  F_{\mu\kappa},\,F_{\nu\lambda}\right\}  -\left\{
A_{\kappa},\,\partial_{\lambda}F_{\mu\nu}+D_{\lambda}F_{\mu\nu}\right\}
\right)  +O(\theta^{2}),\\
\widetilde{\lambda}  &  =\lambda+\frac{i}{4}\theta^{\alpha\beta}\left\{
\partial_{\alpha}\lambda,~A_{\beta}\right\}  +O(\theta^{2}).
\end{align*}
The deformed complex vierbein $\widetilde{L}$ is defined by the relation%
\[
\widetilde{L}\left(  g~L~g^{-1},~g~A~g^{-1}+g~dg^{-1}\right)  =\widetilde
{g}\ast\widetilde{L}\left(  L,A\right)  \ast\widetilde{g}_{\ast}^{-1}.
\]
Its solution is given by
\[
\widetilde{L}_{\mu}=L_{\mu}+\frac{i}{2}\theta^{\kappa\lambda}\left\{
\partial_{\kappa}L_{\mu}+\frac{1}{2}\left[  A_{\kappa},L_{\mu}\right]
,\,\,A_{\lambda}\right\}  +O(\theta^{2}).
\]
The component form of these relations read%
\begin{align*}
\widetilde{a}_{\mu}  &  =a_{\mu}+\frac{1}{2}\theta^{\kappa\lambda}\left(
a_{\kappa}\left(  2\partial_{\lambda}a_{\mu}-\partial_{\mu}a_{\lambda}\right)
+b_{\kappa}\left(  2\partial_{\lambda}b_{\mu}-\partial_{\mu}b_{\lambda
}\right)  \right. \\
&  \quad\left.  +\frac{1}{8}\omega_{\kappa}^{\;ab}\left(  \partial_{\lambda
}\omega_{\mu}^{\;ab}+R_{\lambda\mu}^{\quad ab}\right)  \right)  +O(\theta
^{2}),\\
\widetilde{b}_{\mu}  &  =b_{\mu}+\frac{1}{2}\theta^{\kappa\lambda}\left(
a_{\kappa}\left(  2\partial_{\lambda}b_{\mu}-\partial_{\mu}b_{\lambda}\right)
+b_{\kappa}\left(  2\partial_{\lambda}a_{\mu}-\partial_{\mu}a_{\lambda
}\right)  \right. \\
&  \left.  -\frac{1}{16}\epsilon_{abcd}\;\omega_{\kappa}^{\;ab}\left(
\partial_{\lambda}\omega_{\mu}^{\;cd}+R_{\lambda\mu}^{\quad cd}\right)
\right)  +O(\theta^{2}),\\
\widetilde{\omega}_{\mu}^{\;ab}  &  =\omega_{\mu}^{\;ab}+\frac{1}{2}%
\theta^{\kappa\lambda}\left(  a_{\kappa}\left(  \partial_{\lambda}\omega_{\mu
}^{\;ab}+R_{\lambda\mu}^{\quad ab}\right)  +\omega_{\kappa}^{\;ab}\left(
2\partial_{\lambda}a_{\mu}-\partial_{\mu}a_{\lambda}\right)  \right. \\
&  \quad\left.  +\frac{1}{2}\epsilon_{abcd}\left(  b_{\kappa}\left(
\partial_{\lambda}\omega_{\mu}^{\;cd}+R_{\lambda\mu}^{\quad cd}\right)
+\omega_{\kappa}^{\;cd}\left(  2\partial_{\lambda}b_{\mu}-\partial_{\mu
}b_{\lambda}\right)  \right)  \right)  +O(\theta^{2}),\\
\widetilde{e}_{\mu}^{a}  &  =e_{\mu}^{a}-\theta^{\kappa\lambda}%
\biggl(a_{\lambda}\bigl(\partial_{\kappa}e_{\mu}^{a}+\frac{1}{2}\omega
_{\kappa}^{\;ae}e_{\mu}^{e}\bigr)-\frac{1}{4}\epsilon_{abcd}\omega_{\lambda
}^{\;cd}\bigl(\partial_{\kappa}f_{\mu}^{b}+\frac{1}{2}\omega_{\kappa}%
^{\;be}f_{\mu}^{e}\bigr)\biggr)+O(\theta^{2}),\\
\widetilde{f}_{\mu}^{a}  &  =f_{\mu}^{a}-\theta^{\kappa\lambda}%
\biggl(a_{\lambda}\bigl(\partial_{\kappa}f_{\mu}^{a}+\frac{1}{2}\omega
_{\kappa}^{\;ae}f_{\mu}^{e}\bigr)+\frac{1}{4}\epsilon_{abcd}\omega_{\lambda
}^{\;cd}\bigl(\partial_{\kappa}e_{\mu}^{b}+\frac{1}{2}\omega_{\kappa}%
^{\;be}e_{\mu}^{e}\bigr)\biggr)+O(\theta^{2}).
\end{align*}
As an alternative to the deformed action obtained in this section, one can use
the Seiberg-Witten map for the fields $\widetilde{L}_{\mu}$ and $\widetilde
{A}_{\mu}$ and then substitute the undeformed solution for $\omega_{\mu ab}$
in terms of $e_{\mu}^{a}$ and $f_{\mu}^{a}$ . The resulting expressions would
be very complicated which shows that the use of the SW map in obtaining the
deformed action is not practical for the gravitational system. These
expressions might simplify for specific solutions where $\omega_{\mu ab}$,
$e_{\mu}^{a}$ and $f_{\mu}^{a}$ are given.

\section{Conclusions}

The idea that the gravitational field could be complex is not new and was
first considered by Einstein and Stauss \cite{Einstein} motivated by the
unification of electromagnetism with gravity. The work of Weyl \cite{weyl} and
Cartan \cite{cartan} on spinors in general relativity and of Utiyama
\cite{utiyama} and Kibble \cite{kibble} relating gravity to a gauge theory of
the Lorentz group, showed how general relativity could be formulated based on
the $SL(2,\mathbb{C})$ gauge invariance \cite{salam}. This symmetry also
played a crucial part in determining Ashtekar variables \cite{sen},
\cite{ashtekhar}. The $SL(2,\mathbb{C})$ symmetry acts as a gauge symmetry of
the spin-connection, and in a first order formalism gives the correct kinetic
terms for the vierbein. It is also possible to include torsion in the
spin-connection to accommodate the antisymmetric $B$ field appearing in string
theory and give it a kinetic term. In this paper we have shown that it is
possible to go further and complexify the vierbein, keeping the
$SL(2,\mathbb{C})$ symmetry. We have proposed an action with the exceptional
property that when the spin-connection, which appears quadratically, is
eliminated by its equation of motion, then both the real and imaginary parts
of the metric propagate. One combination protected by diffeomorphism
invariance will produce the massless graviton, while the other will produce a
massive graviton coupled to a scalar field. This is identical to the spectrum
of bigravity, but has the advantage of using a minimal number of fields. We
have worked out only the linearized approximation of the theory and shown that
all fields acquire the correct kinetic terms. The computation is not simple,
but it is very important to go one step further and determine the higher order
interactions. Such calculation can only be performed perturbatively because
the massless and massive gravitons are linear combinations of the real and
imaginary parts of the complex vierbein $L$ and these tensor combinations
should be inverted. It would be very enlightening to find some special
solutions for this theory which are generalizations of the Schwarzschild and
de Sitter solutions.

When coordinates do not commute and fields are defined on such noncommutative
space, ordinary products must be replaced with star products. Commutators of
Lie algebra valued fields using star products, would result in both
commutators and anticommutators in terms of the undeformed fields. This makes
it necessary to extend the gauge group form $SL(2,\mathbb{C})$ to
$GL(2,\mathbb{C}).$ Having the proposed action for complex gravity based on
the requirement that all terms must be four-forms, the extension carries
through without any complications by replacing ordinary products with star
products. It is then a straightforward matter to determine the deformed action
to second order in the deformation parameter $\theta^{\mu\nu}.$ We have only
touched the surface in this direction, and many questions remain to be
addressed such as the effect of the deformed parameters on quantization of the
theory, finding the SW map of some specific solutions, and generalization to
non-constant parameters $\theta^{\mu\nu}.$ These questions and others will
hopefully be addressed in future investigations.

\end{document}